# Comparison of Different Machine Learning Approaches to Predict Viscosity of Tri-n-Butyl Phosphate Mixtures Using Experimental Data

**Faranak Hatami [1] and Mousa Moradi [2],\***

[1] Department of Physics and Applied Physics, University of Massachusetts, Lowell, MA 01854, USA; faranak_hatami@uml.edu

[2] Department of Biomedical Engineering, University of Massachusetts, Amherst, MA 01003, USA; mousamoradi@umass.edu

\* Correspondence: mousamoradi@umass.edu

**Abstract:** Tri-n-butyl phosphate (TBP) is a solvent that is commonly used in a variety of industries, including the nuclear and chemical industries, for its ability to dissolve and purify various inorganic acids and metals. It is often used in hydrometallurgical processes to separate and purify these substances. Machine learning models offer a promising alternative to traditional methods for predicting the viscosity of TBP mixtures. By training machine learning models on a dataset of viscosity measurements, it is possible to accurately predict the viscosity of TBP mixtures at different compositions, densities, and temperatures, which can save time and resources and reduce the risk of exposure to toxic solvents. This paper aimed at proposing Machine Learning (ML) techniques to automatically predict the viscosity of TBP mixtures using experimental data. For comparison peruses, we trained five different ML algorithms including Support Vector Regressor (SVR), Random Forest (RF), Logistic Regression (LR), Gradient Boosted Decision Trees (XGBoost), and Neural Network (NN). We collected a total of 511 measurements for TBP mixtures with temperature-based density, at different compositions, containing hexane, dodecane, cyclohexane, n-heptane, toluene, and ethylbenzene measured at temperatures of T= (288.15, 293.15, 298.15, 303.15, 308.15, 313.15, 318.15, 323.15, and 328.15) K. The results revealed that the NN model with 25 and 50 neurons in the hidden layers could achieve the best viscosity predictions for a system of TBP mixtures. The NN model outperformed other regular ML models in terms of Mean Square Error (MSE) of 0.157 % and adjusted $R^2$ of 99.72 % on the test data set. This paper demonstrated that the NN model can be an appropriate option to accurately predict the viscosity of TBP + Ethylbenzene with a margin of deviation as low as 0.049 %.

## 1. Introduction

Reducing the volume and the level of radioactivity of high-level wastes before their disposal, as well as recovering unused uranium, along with plutonium has been widely investigated. The first successful solvent extraction process for the recovery of pure uranium and plutonium was developed as the PUREX process (Plutonium and Uranium Recovery by Extraction [1]. Tri-n-butyl phosphate (TBP) is one of the most important candidates used as an extractant of uranium, plutonium, zirconium, and other metals liquid-liquid in atomic energy, hydrometallurgical and chemical industries. For solvent extractions a polar or nonpolar diluent is employed with TBP, changing the density making phase separation easier, and lowering viscosity hence improving flow and kinetic properties [1-7]. Thus, TBP and its complexes with organic solutes are viscous and have densities relatively close to that of the aqueous phase in typical liquid-liquid solvent extraction systems, in which viscosity is a very important transfer property in the scale-up of liquid applications [8].

Due to its importance, researchers have accomplished some efforts to determine the viscosities of various TBP−dilution systems experimentally. In 2007, Tian et al. [9] measured the densities and viscosities of the binary mixtures of TBP with hexane and dodecane systems. Following this work, in 2008, Fang et al. [10] measured these experimental values using cyclohexane and n-heptane with binary mixtures of TBP at atmospheric pressure in the whole composition range and over different temperatures. Fang reported that when the Grunberg Nissan equation was used to measure viscosity, the experimental data were measured with less than 1% average absolute deviation (ADD) at temperatures (from 288 to 308) K. Sagdeev et al. [11]

in 2013, measured the density and viscosity of n-heptane over a wider temperature range from 298 K to 470 K using the hydrostatic weighing and falling-body techniques. They calculated the ADD of 0.21% to 0.23% between the present and reported measurements of the density for n-heptane. In the same year, Basu et al. [2] obtained the partial molar volumes of the respective components using a binary mixture of TBP with hexane, heptane, and cyclohexane at different temperatures from (298.15 to 323.15) K. Basu reported that a binary mixture of TBP with hexane could achieve a consistent negative deviation from ideality. Most recently, in 2018, Billah et al. [3] measured densities, viscosities, and refractive indices, for the binary mixtures of TBP under atmospheric pressure with toluene and ethylbenzene over the entire range of composition, and at different temperatures from T = (303.15 to 323.15) K.

Artificial intelligence and Machine Learning (ML) have been widely used in all areas of computational science including biomedical sciences, drug delivery, chemistry, and material sciences to predict future behavior of systems based on data history [12-17]. However, in the field of computational chemistry, most of the conventional TBP mixture systems to measure the viscosity only focused on the viscosity estimation of pure liquids and binary mixtures. Therefore, prediction models for the viscosity of TBP mixtures are needed in the practical process calculation and equipment design. Despite the wide applications and popularity, the conventional TBP mixture systems to measure the viscosity, besides having a time-consuming and tedious process, required a multistage preparation method which might be harmful to human health and the environment due to toxic organic solvents. For this reason, machine learning-based models that can give reliable predictions on the viscosities of liquids are crucial.

The literature survey revealed that, up to now, there is no study on modeling the viscosity of TBP mixtures using ML techniques. This gap in the research inspired us to propose ML models to accurately predict the viscosity of TBP diluted in solvents using experimental data obtained from the literature [2, 3, 9, 10]. In the present study, for the first time, an accurate and efficient artificial neural network has been proposed to predict the relative viscosity of TBP. This paper also demonstrates that ML models can predict viscosity of TBP mixtures with high degree of accuracy. We show that among five ML models developed in this study, the NN model can be a suitable option to accurately predict the viscosity of TBP mixtures with a lower Margin of Deviation (MOD) compared to other regular ML models. To have a fair comparison, we developed five different ML algorithms including LR, SVR, RF, XGBoost and NN on the same experimental data points (4599 = 511×9) for the viscosity of TBP mixtures at different compositions, density, and temperatures. We hypothesized that ML models can provide accurate predictions of viscosities compared with the actual (observed) values. To test this hypothesis, a two-sample t-test will be performed on the predicted viscosity values.

## 2. Materials and Methods

The general block diagram of the proposed algorithm to predict the dynamic viscosity of TBP mixtures is shown in Figure 1. The dataset consists of 511 measurements for TBP mixtures with temperature-based density, at different compositions. We pre-processed the data by standardizing it to have the same scale for all features and, the dimensionality of the data was reduced to have the most efficient training. More details of the other blocks in Figure 5 will be provided in the following sections.

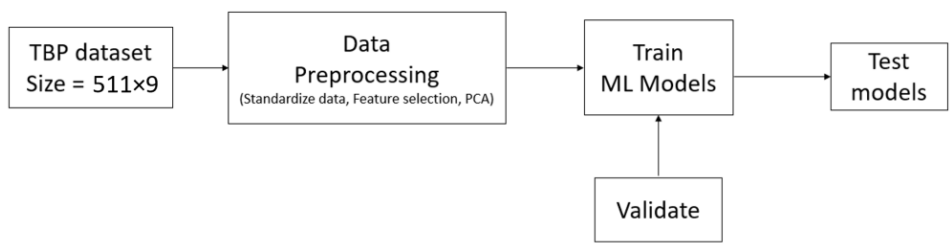

Figure 1. The main steps used in this study are to predict dynamic viscosity of TBP mixtures.

*2.1. Experimental Database and Feature Visualization*

The databases of this study were formed from results reported in the literature. A total of 511 experimental points of TBP mixtures with temperature-based density, at different compositions, containing hexane, dodecane, cyclohexane, n-heptane, toluene, and ethylbenzene are collected from various scientific publications [3, 9, 10]. The original dataset has nine descriptors (features) such as temperature, density, and seven mentioned compositions to estimate the viscosity. In Figure 2 (a-h), the x-axis indicates the values in the dataset and the y-axis shows the Frequency of each value and Figure 2 (i) represents the viscosity changes in the dataset following a normal distribution with N (1.47, 0.81).

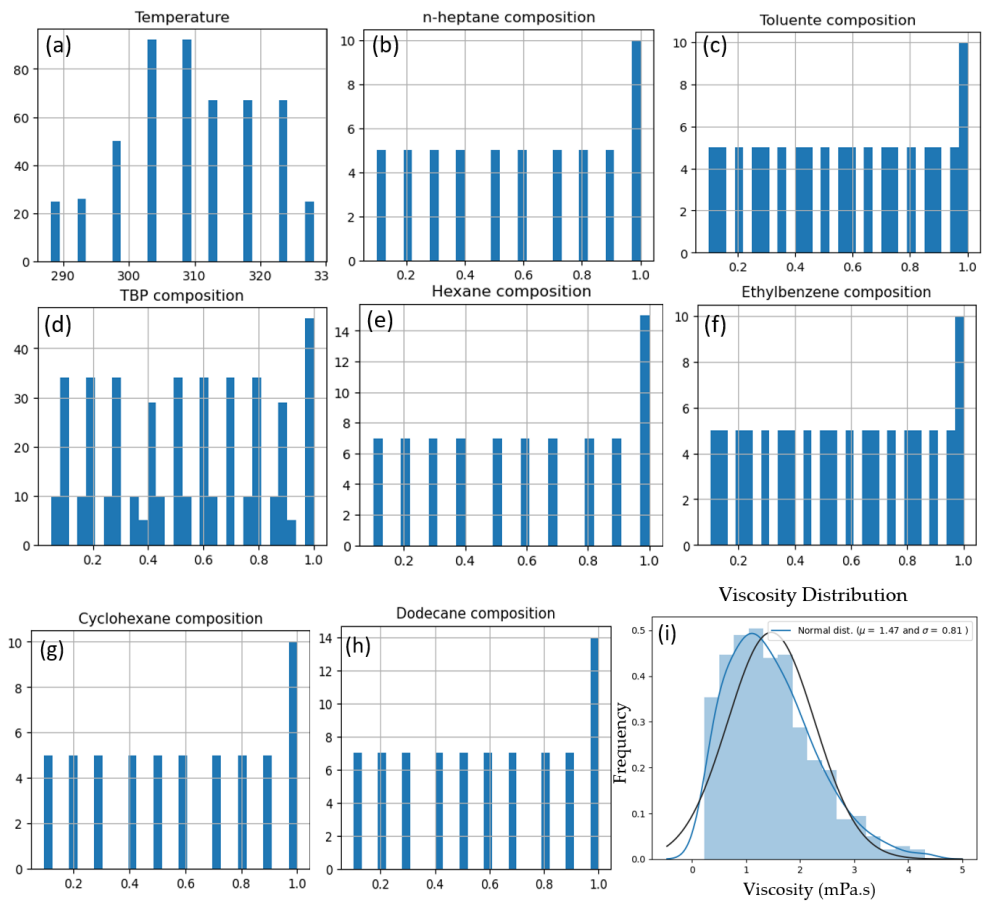

Figure 2. Distribution of features used in this study showed in (a-h). Viscosity has a normal distribution with N (1.47, 0.81) as shown in (i).

## 2.2. Correlation Matrix

To find how the features were related to one another or to the target variable (viscosity), the Pearson product-moment correlation coefficient (often abbreviated as Pearson's r) was used [18]. The correlation heatmap in Figure 3 shows the linear dependence between pairs of features. Two features have a perfect positive correlation if r is equal to the maximum positive value (r = 1.0), no correlation if r = 0, and a perfect negative correlation if r is equal to the negative value (r = -1.0). To fit a linear regression model, we were interested in those features that have a high correlation with our target variable (viscosity). Looking at the correlation matrix, TBP composition has the largest positive correlation with viscosity (r = 0.8). Inversely, the temperature shows the largest negative correlation with viscosity (r = - 0.5).

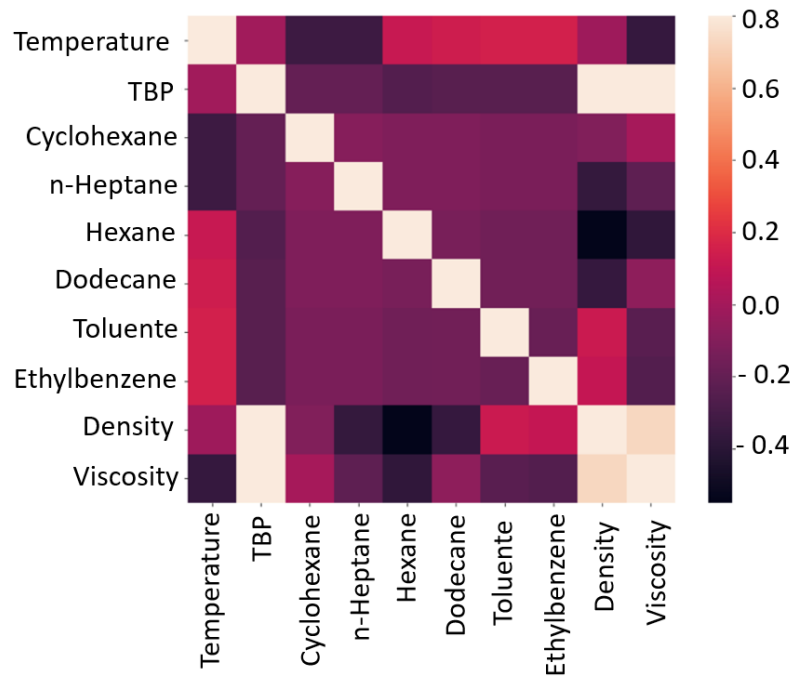

Figure 3. Correlation heatmap matrix for input features and the output target.

The negative correlation between viscosity and temperature was clearly visible in Figure 4 (a). A three-dimensional plot of 511 experimental data for the relative viscosity of the mixture covering a mole fraction ranging from 0 to 1% of TBP composition and temperature ranging from 290 to 330 K has been shown in Figure 4 (b). As it is shown, variations in viscosity due to changes in temperature and the effect of mole fraction of TBP on viscosity are noticeable. These figures indicated that the increased mole fraction of TBP, the more rapid changes occur, and the more considerable changes.

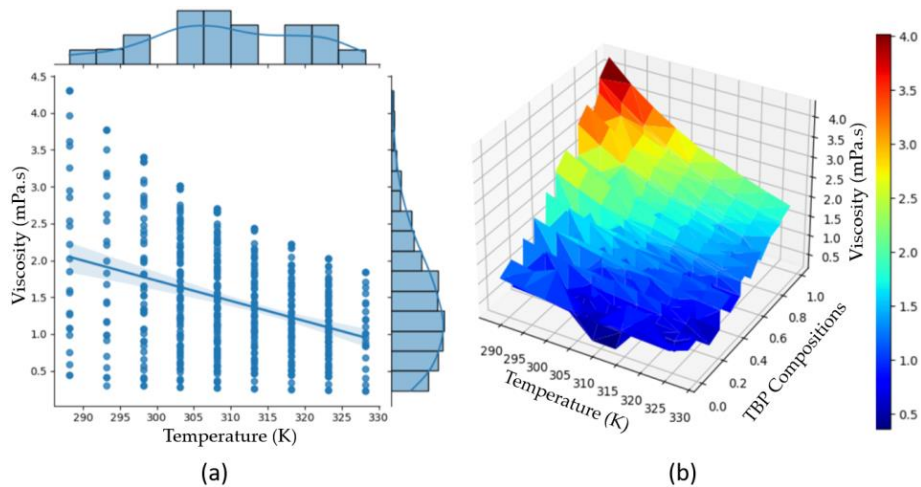

Figure 4. (a) Association of temperature with viscosity. (b) Relative viscosity versus TBP concentration (r = 0.8) and temperature (r = - 0.5).

*2.3. Feature Selection and Principal Component Analysis*

Principal Components Analysis (PCA) is an unsupervised outstanding dimensionality feature-reduction technique, and it has been widely applied to datasets in all scientific domains [19]. In other words, PCA tends to map data points from a high to a low dimensional space while keeping all the relevant linear structures intact. The strong optimality properties for the resulting low-dimensional embedding arise from a precise mathematical framework that underlies PCA. To select the most informative features, we used pairwise correlation in python and set Pearson correlation with 5% - 95% to select highly correlated features. We applied PCA to correlated features to reduce dimensionality to eight principal components which explained more than 99% of data variability. Figure 5 (a) indicates the explained variance for all the features and the variation distribution of selected important features has been shown in Figure 5 (b), disregarding density to train the ML algorithms. This shows that temperature and n-heptane composition have the highest and smallest variation, respectively, compared with other features.

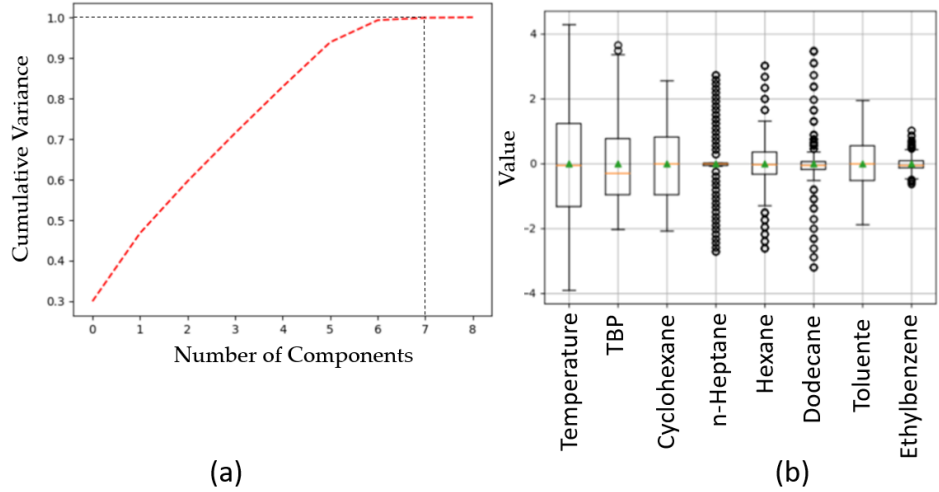

(a)  (b)

Figure 5. Explained variance (a), eight first features explained almost all variability, (b) distribution of important features.

To be more precise, temperature, and the solvents compositions ($x_i$) of each species were considered as input variables. The functionality of viscosity of solvents are given as bellow:

$$\mu = f(T, X_1, X_2, X_3, X_4, X_5, X_6, X_7) \qquad (1)$$

Where $X_1$ to $X_7$ are the solvents and T = (288.15, 293.15, 298.15, 303.15, 308.15, 313.15, 318.15, 323.15, and 328.15) K are experimental temperature used to measure compositions. After identifying and collecting the data set, the next step was to select ML models to predict the viscosity. To do so, for comparison purposes, we trained five different ML algorithms including Support vector regressor (SVR), RF, LR, XGB and NN, which will be explained in the following section. To train these models, we randomly split our dataset to 80% for training and 20% for validation and testing. In this study, the training set was used to generate the ML model and the test set was utilized to investigate the prediction capability and to validate the trained model.

### 2.4. Machine Learning Models

All models discussed in this section were built in Python 3.8 with Scikit-Learn 1.1.3, Keras 2.9, TensorFlow 2.9.1 libraries trained by an RTX 3090 GPU with 24 GB memory, and CUDA 11.4.

### 2.4.1. Support Vector Regression (SVR)

SVR algorithm formulates function approximation problem as an optimization problem, while minimizing the distance between the predicted and the desired outputs [20-22]. Because our problem was not linearly separable in input space, a kernel was used to transform the data to a higher-dimensional space, referred to as kernel space, where data will be linearly separable. In this study, the Scikit-learn [23] (version 1.1.3) library with SVR class with parameters including a radial basis function ('RBF) kernel with a 0.001 tolerance for stopping criterion, and the regularization parameter C=1 have been used. We performed a linear SVR which is expressed below:

$$f(x) = \omega^T X + b \qquad (2)$$

Where X, ω = (ω₁, ω₂, …, ωₙ) ∈ $R_n$, b ∈ R and T are respectively the input or support vectors, the weight vector, the intercept, and transpose operator. The optimization problem for training the linear SVR is given by:

$$Min \left( \frac{1}{2} ||\omega||^2 + \frac{C}{2}\sum_{i=1}^{n} e_i^2 \right) \quad (3)$$

Where C is a positive regularization parameter (C=1), the penalty for incorrectly estimating the output associated with input vectors.

*2.4.2. Gradient Boosted Decision Trees (XGBoost)*

XGBoost regressor in this study, combined 100 decision trees to reduce the risk of overfitting that each individual tree faces. XGBoost used a method called boosting, which combined weak learners sequentially, so that each new tree corrected the errors of the previous one [24, 25]. In this work, the XGBoost model developed with a learning rate of 0.3 and the number of estimators = 100.

*2.4.3. Random Forest (RF)*

By employing distinct subsets of accessible characteristics, many independent decision trees can be constructed simultaneously on different segments of the training samples. Bootstrapping guarantees that any decision tree inside the random forest is distinct, causing the lowering of RF variance [26]. We found 100 trees in the forest to be optimum in our RF model.

*2.4.4. Logistic Regression (LR)*

The main advantage of LR is to avoid confounding effects by analyzing the association of all variables together [27, 28]. In this work, we used "L2" regularization method which is a ridge regression and can add a squared magnitude of coefficients as the penalty term to the loss function regularization.

*2.4.5. Neural Network Architecture*

In the NN architecture, we utilized a feed-forward back propagation approach to update the weights and to find the best map to relate input features to the target output [29, 30]. The structure of the feed forward network used in this work has been shown in Figure 6, containing two hidden layers with 25 and 50 neurons for the first and second hidden layers, respectively. The 'ReLu' activation function was used to pass only positive weights as defined in Equation (4). We used a cross validated approach using GridSearchCV method with 5-fold cross validation in python to select the optimal parameters [31, 32].

$$f(x) = max\,(0, x) \quad (4)$$

*2.5. Performance Metrics*

The performance of the trained models was measured using three parameters as Mean Absolute Error (MAE), Mean Square Errors (MSE) and adjusted correlation coefficient ($R^2$). Regression analysis has been applied to assess the network capability for ternary viscosity prediction. The coefficient of determination has been used as a measure to evaluate how the trained model was correlated to the experimental data. The performance metrics used in this study were defined as below:

$$MAE = \frac{1}{N}\sum_{i}^{N} |\mu_i^{obs} - \mu_i^{pred}| \quad (5)$$

$$MSE = \frac{1}{N}\sum_{i}^{N} (\mu_i^{obs} - \mu_i^{pred})^2 \quad (6)$$

$$R^2 = \frac{\sum_{i}^{N}(\mu_i^{obs}-\overline{\mu_i})^2 - \sum_{i}^{N}(\mu_i^{obs}-\mu_i^{pred})^2}{\sum_{i}^{N}(\mu_i^{obs}-\overline{\mu_i})^2} \quad (7)$$

$$R_{Adjusted}^2 = 1 - (1-R^2)\frac{N-1}{N-p-1} \quad (8)$$

Where N is the number of viscosity data points, $\mu^{obs}$ is the $i_{th}$ observed value of the viscosity, $\mu^{pred}$ is the predicted viscosity with the ML model, $\bar{\mu}$ is the average value of the experimental viscosity data, and p is the number of predictors.

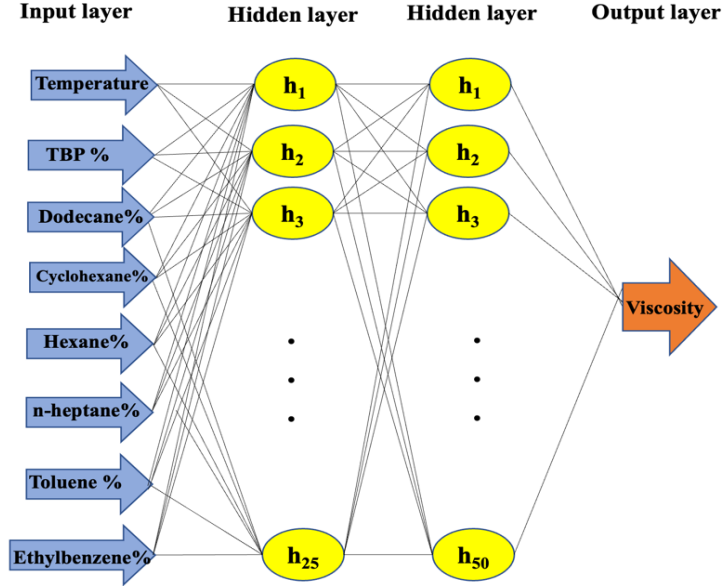

Figure 6. The optimal structure of neural network developed in this study.

*2.6. Statistical Analysis*

The statistical analysis was performed in RStudio Version 2022.07.0+548. To determine whether there are any statistically significant differences between observed and predicted viscosity values, for each ML model, it was assumed that feature distribution was normal, and samples were taken from independent features. Then for each model, we checked the null hypothesis was assessed by calculating mean μ for viscosity values to test whether $\mu_{observed} = \mu_{predicted}$.

The Margin of Deviation (MOD) was used to calculate the error bars as defined below:

$$MOD = \frac{\mu^{obs} - \mu^{pred}}{\mu^{obs}} \times 100 \qquad (9)$$

**3. Results**

In this section, the performance of the different approaches and optimum parameters for the developed ML algorithms is discussed. To evaluate the performance of ML models in predicting the viscosity of TBP mixtures, we developed five different ML models from scratch including Logistic Regression (LR), Support Vector Regressor (SVR), Random Forest (RF), Gradient boosted decision trees (XGBoost), and Neural Network (NN). The developed models trained on 408 observations with 8 features (3264 total data points) and tested on 103 observations with 8 features (824 total data points) at T= (288.15 to 328.15) K. The learning curves for the NN model are shown in Figure 7. The NN model started to converge after 20 epochs. Table 1 compares the performance of all developed models in terms of adjusted $R^2$, Mean Square Error (MSE) and Mean Absolute Error (MAE). As shown in Table 1, the NN model was able to achieve the highest performance at 0.157 % MSE and 99.97 % adjusted $R^2$. Following the NN, ranking from the top, XGBoost, RF, SVR, and Logistic Regression could achieve adjusted $R^2$ of (99.54, 99.22, 99.06, and 93.10) % on the test dataset, respectively.

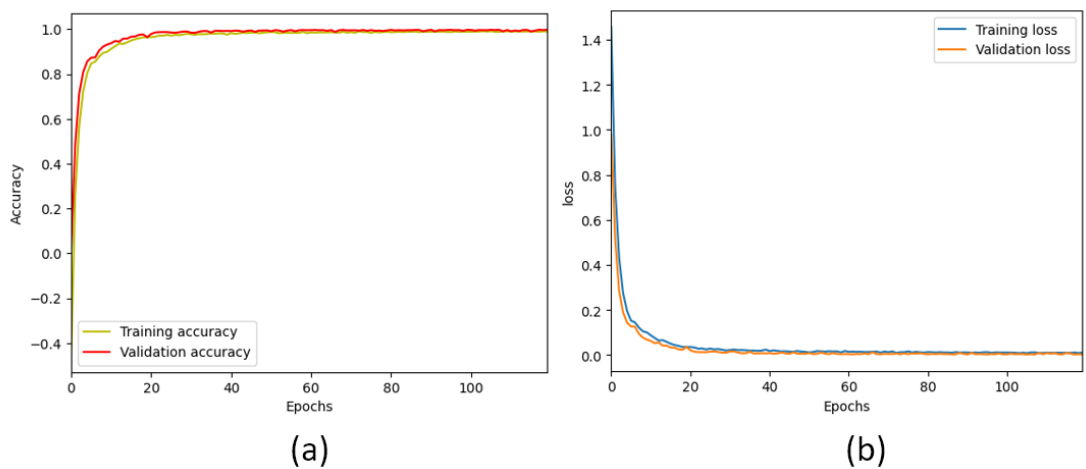

Figure 7. Learning curves for the NN for 120 Epochs: (a) Accuracy profile; (b) Loss profile. The model was converged after 20 epochs. 10% dropout was used to avoid overfitting.

Table 1. Comparison of five ML models in predicting the viscosity of TBP mixtures on the test set.

| ML model | Adj. $R^{2*}$ (%) | MAE (%) | MSE (%) |
|---|---|---|---|
| NN | 99.72 | 2.860 | 0.157 |
| XGBoost | 99.54 | 3.751 | 0.315 |
| RF | 99.22 | 5.372 | 0.530 |
| SVR | 99.06 | 5.523 | 0.650 |
| LR | 93.16 | 1.565 | 0.474 |

*Adjusted R-squared.

The predicted viscosities were compared with the observed values obtained from experiments using scatter and box plots as shown in Figure 8. Training and testing of all models proved that the NN and the LR models were the best and the worst models, respectively, in predicting the viscosity of TBP mixtures (First and second columns in Figure 8). Statistical analysis showed, for the NN model, we had very strong evidence that predicted and observed viscosities were similar on the train set (P < 1e-14) although it was marginally significant on the test set (P = 0.055). Similarly, for the LR model, the predictions and observed values were different on both train and test sets (P = 0.99 and P = 0.64).

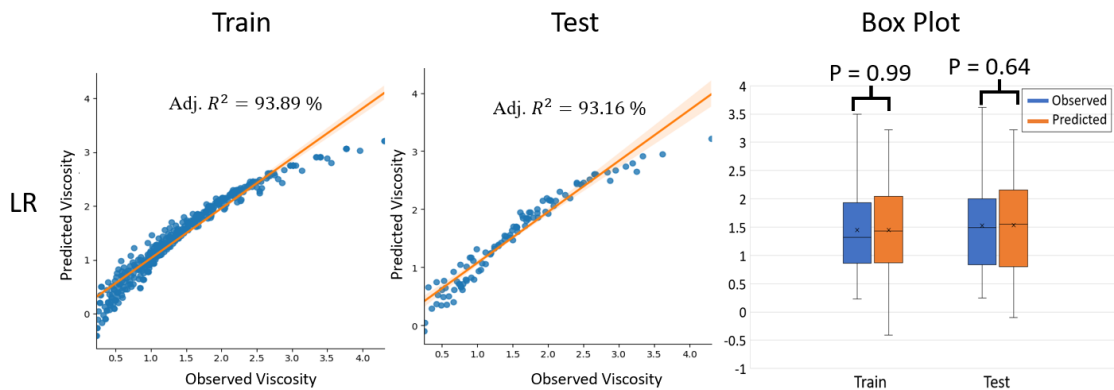

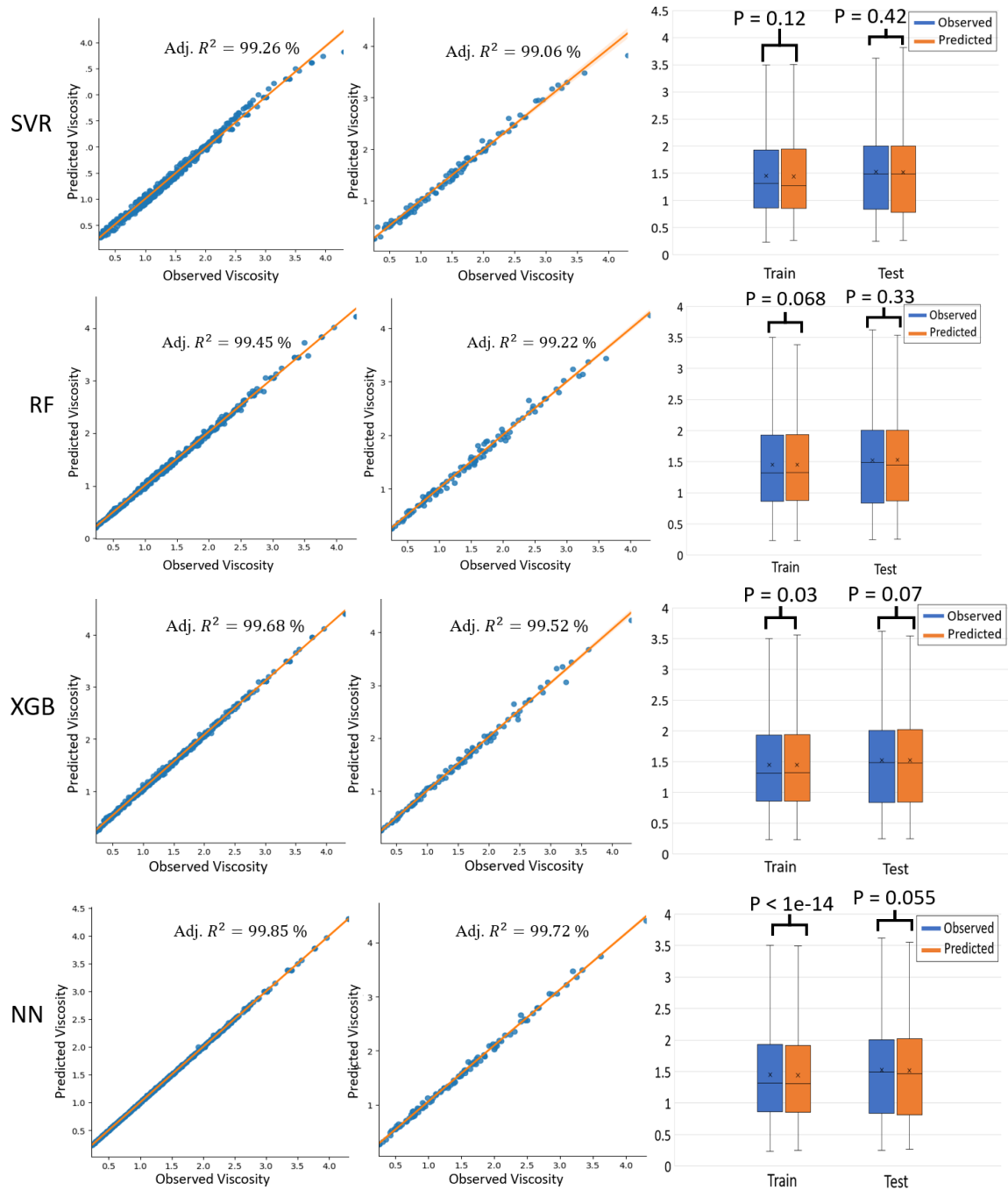

Figure 8. Evaluating five developed models in predicting viscosity of TBP mixtures. The labels of 'LR', 'SVR', 'RF','XGB', and 'NN' refer to the respective five image rows and "Train", "Test", and" Box Plot" refer to the respective three columns. 95% confidence interval was used to plot the data. 0.05 significance level was used to compare predicted and observed values.

Figure 9 presented the NN performance using the top four features correlated with viscosity. A wide range of temperatures and mole fractions of three types of composition (Cyclohexane, n-Heptane, and Ethylbenzene) are used to train the model. The proposed NN model could predict viscosities with a low MOD for the given features. The larger the MOD, the more uncertainty in viscosity predictions. The MOD of viscosity when TBP + Cyclohexane was used as a feature was 0.27 % which reveals the NN had lower certainty in predicting viscosities when TBP + Cyclohexane was used as an input (Figure 9 (a)). On the other hand, TBP + Ethylbenzene provided the highest certainty in predicting viscosity with 0.049 % MOD (Figure 9 (b)).

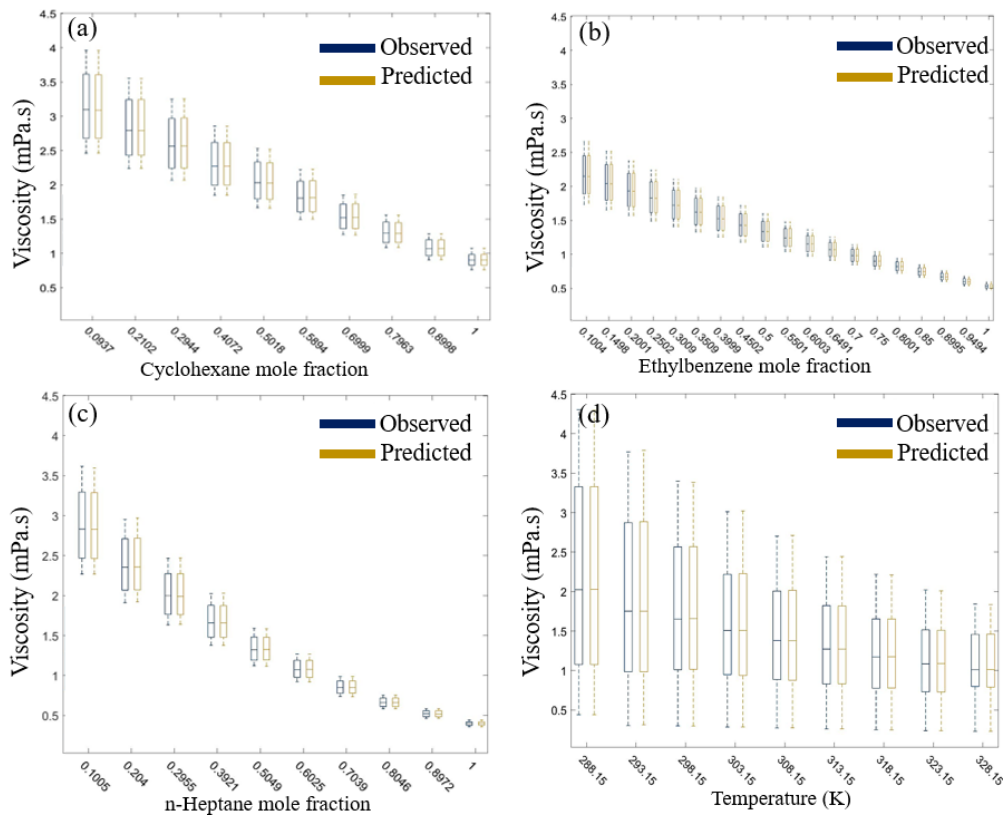

Figure 9. Comparison between observed and predicted values for four top features associated with NN performance. TBP + Ethylbenzene showed the best NN performance in predicting viscosity.

Figure 10 shows the MOD of all eight features used to train the NN model. As we can see, TBP + Temperature, TBP + Hexane, and TBP + Dodecane had a negative MOD which means the predicted viscosities using these features were larger than the actual values. Vice versa, mixtures of TBP with Cyclohexane, n-Heptane, Ethylbenzene, Toluente had a positive MOD which represents the predicted viscosity values were not greater than their actual values. We can observe from Figure 4 that a mixture of TBP + Cyclohexane can provide the largest positive MOD of 0.16 % while temperature indicated the largest negative MOD of - 0.12 %.

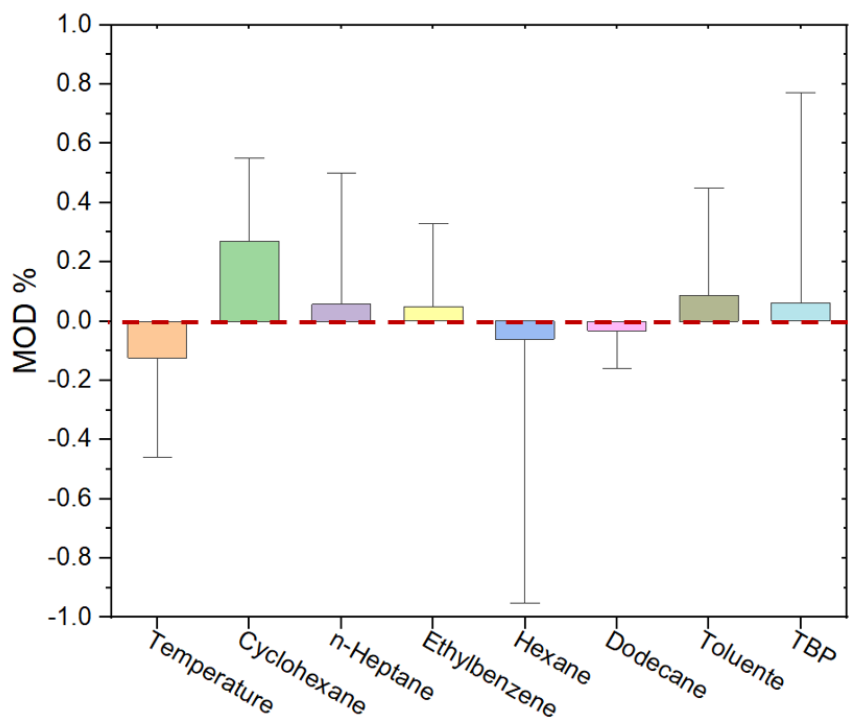

Figure 10. The margin of deviation of NN results with respect to the type of solvent (feature).

4. **Discussion**

This paper aimed at predicting viscosity of TBP mixtures using five ML models including LR, SVR, RF, XGBoost, and NN. To have a fair comparison, all models trained on 408 observations from 8 features as TBP, hexane, dodecane, cyclohexane, n-heptane, toluene, and ethylbenzene measured at temperatures of T= (288.15, 293.15, 298.15, 303.15, 308.15, 313.15, 318.15,323.15, and 328.15) K.

To have a fair comparison, we developed all ML models on the same experimental data points (4599 = 511×9). The NN model trained on 408 observations resulted in 0.059% MSE and 99.85 % adjusted $R^2$ while it could achieve 0.157 % MSE and 99.72 % adjusted $R^2$ on the test set. Comparatively, LR achieved the lowest certainty in predicting viscosity with the adjusted $R^2$ of 93.16 % on the test dataset. We reported that the proposed NN with 25 neurons in the first hidden layer and 50 neurons in the second hidden layer is the optimum model in viscosity predictions for TBP mixtures. The proposed NN model could achieve the highest certainty for a system of TBP + Ethylbenzene with 0.049 % MOD.

The main limitation of this study was the small dataset size. Although we fined-tune all ML models to avoid overfitting and underfitting using the GridSearchCV technique and randomly dropping out 10 % of the training data, a larger dataset is desired to get more accurate results and achieve lower errors. Future work will be to expand experimental data to include more compositions that are associated with viscosity predictions. Also, in this study, we did not predict density as we considered it as one of the input features, but future work can be focused on predicting density along with the viscosity of TBP mixtures.

5. **Conclusions**

Artificial intelligence and machine learning have been widely used in all areas of computational science including biomedical sciences, drug delivery, chemistry, and material sciences to predict future behavior of systems based on data history. However, in the field of computational chemistry, most of the conventional TBP mixture systems to measure the viscosity only focused on the viscosity estimation of pure liquids and binary mixtures. On the other hand, there is not enough study for the viscosity prediction of mixture systems rather than binary mixtures.

We collected experimental data on TBP mixtures containing hexane, dodecane, cyclohexane, n-heptane, toluene, and ethylbenzene at various temperatures, and used this data to train five different machine learning algorithms: support vector regressor, random forest, logistic regression, gradient boosted decision trees, and neural network. We found that the neural network model was the most accurate at predicting the viscosity of TBP mixtures, with a mean square error of 0.157% and an adjusted $R^2$ of 99.72% on the test data set. The neural network model was also able to accurately predict the viscosity of TBP + ethylbenzene mixtures with a margin of deviation as low as 0.049%. Overall, the results of this study suggest that machine learning techniques can be used effectively to predict the viscosity of TBP mixtures.

**Author Contributions:** Conceptualization, methodology, data preparation and cleaning, writing, original draft preparation F.H.; formal analysis, review and editing, visualization, project administration, M.M.

**Funding:** This research received no external funding.

**Institutional Review Board Statement**: Not applicable.

**Informed Consent Statement**: Not applicable.

**Data Availability Statement**: The data and models presented in this study are openly available at the author's GitHub page: https://github.com/faranak1991?tab=repositories

**Conflicts of Interest**: The authors declare no conflict of interest.